# Atomic Resolution Imaging of CrBr$_3$ using Adhesion-Enhanced Grids

*Matthew J. Hamer[1,2], David G. Hopkinson[2,3], Nick Clark[2,3], Mingwei Zhou[1,2], Wendong Wang[1,2] Yichao Zou[3], Daniel J. Kelly[2,3], Thomas H. Bointon[2], Sarah J. Haigh[2,3,\**], Roman V. Gorbachev[1,2,4,\*]*

[1]Department of Physics and Astronomy, University of Manchester, Oxford Road, Manchester, M13 9PL, UK

[2]National Graphene Institute, University of Manchester, Oxford Road, Manchester, M13 9PL, UK

[3]Department of Materials, University of Manchester, Oxford Road, Manchester, M13 9PL, UK

[4]Henry Royce Institute, Oxford Road, Manchester, M13 9PL, UK

*Roman@Manchester.ac.uk, **Sarah.Haigh@Manchester.ac.uk

**Abstract**

Suspended specimens of 2D crystals and their heterostructures are required for a range of studies including transmission electron microscopy (TEM), optical transmission experiments and nanomechanical testing. However, investigating the properties of laterally small 2D crystal specimens, including twisted bilayers and air sensitive materials, has been held back by the difficulty of fabricating the necessary clean suspended samples. Here we present a scalable solution which allows clean free-standing specimens to be realized with 100% yield by dry-stamping atomically thin 2D stacks onto a specially developed adhesion-enhanced support grid. Using this new capability, we demonstrate atomic resolution imaging of defect structures in atomically thin $CrBr_3$, a novel magnetic material which degrades in ambient conditions.

**Introduction**

The family of 2D materials now includes over a hundred members, many of which have been isolated in monolayer form by mechanical exfoliation[1,2], direct chemical growth[3,4] or solution-phase processing[5]. They possess a diverse range of electronic and optical properties, which can vary even for a single material depending upon thickness[6,7] and stacking order[8,9]. Most recently, 2D crystals with magnetic order have been identified, opening a wide range of possibilities for both fundamental research and applications[10]. To date, several of these magnetic van der Waals' materials have been isolated and characterized, including $CrBr_3$[11], $CrI_3$[12], $CrGeTe_3$[13], $MnPS_3$[14] and $FePS_3$[15]. These atomically thin crystals can exhibit both ferromagnetic and antiferromagnetic behavior with strong layer-dependence[16] making them ideal as magnetic components within 2D heterostructure-based devices. For example, magnon-assisted electron scattering has been observed in CrBr$_3$ tunnel barriers with graphene electrical contacts, allowing scope for spin injection into graphene[17]. One of the chief reasons for the relatively late emergence of these materials is their rapid structural and property degradation in ambient conditions, with all of the aforementioned crystals being sensitive to air, light and/or moisture. For example, CrI$_3$ is found to degrade quickly in ambient environments via photocatalytic substitution of iodine by water[18]. This necessitates a complex sample preparation route where samples are exfoliated in an inert gas environment and encapsulated with impermeable 2D materials in order to mitigate degradation[18,19]. Unfortunately, such processing generates additional sample preparation challenges that have hindered characterization, including the use of transmission electron microscopy (TEM). Such TEM investigations would be highly desirable in order to study local defect structures including atomic edge reconstruction, grain boundaries, local ordering, adatom doping, and vacancy dynamics. The key hurdle for TEM experiments is producing large, flat and contamination-free suspended few-layer crystals without exposure to high temperatures or multiple cleaning steps which are liable to damage the material or break the encapsulation barrier[20]. This problem is common to other characterization routes that require suspended 2D crystals, including nanomechanical testing[21-23] and transmission optical measurements[24]. Here we outline a route to reliably fabricate clean free-standing specimens for air-sensitive 2D crystals with nearly 100% yield. This is achieved using a new adhesion-enhanced support grid (AEG), which consists of a Si/SiN$_x$ substrate with added Cr, Au, and MoS$_2$ layers and which has the potential to be produced on a large-scale production *via* wafer scale processing. The ultra-flat MoS$_2$ surface provides improved van der Waals interaction such that the grids are compatible with the widely acclaimed dry-stamp transfer method[19] for 2D heterostructure flake transfer. We demonstrate this new approach by providing the first atomic resolution TEM studies of degradation, defects and stacking in the recently isolated, highly air-sensitive, magnetic 2D crystal, $CrBr_3$.

**Main Body**

Plan view TEM imaging of atomically thin crystals is achieved by suspending a specimen on a perforated support membrane. These support grids can comprise of $Si/SiN_x$, $Si/SiO_2$ or of a (holey) amorphous carbon membrane suspended over a metal mesh[25,26]. The choice of grid and method used to obtain a free standing 2D crystal depends upon the experimental goals, the specimen and the local skill set. TEM specimens of liquid phase exfoliated crystals can be simply drop cast onto conventional support grids, and here the challenge is hunting for suitable areas of free-standing crystal and removing undesirable solvents that may hinder TEM investigations[27]. However, for many exfoliated and grown 2D crystals, as well as all fabricated 2D heterostructures, the specimen is usually prepared using the wet transfer approach, which involves using a micromanipulation system to transfer a 2D crystal, attached to a polymer support layer, directly onto the TEM grid[28]. To achieve high resolution imaging the polymer layer must be dissolved, leaving just the specimen suspended over the holes of the support grid[28]. This technique has a number of problems: (1) poor adhesion between 2D crystals and the support grid often results in failed transfers as specimens delaminate or scroll (see **Supporting Information S2**), (2) solvent surface tension during the drying process often results in the rupturing of suspended regions[29,30] and (3) incomplete removal of the polymer support layer frequently results in a 1-10 nm thick layer of surface contamination which prevents ultimate resolution TEM imaging[31,32]. Contamination can often be removed with high temperature heat treatments or subsequent washing steps; but these are not suitable for all specimens, provide additional points of failure and can degrade the specimen composition. Taken together these factors seriously limit the yield for successfully fabricating free-standing 2D crystals, wasting many thousands of research hours depending upon the type and size of the specimen (see **Table 1**). A more reliable method for specimen preparation is thus required to facilitate efficient TEM studies of many new 2D materials and their heterostructures, as well as for a broad range of optical and nanomechanical techniques where clean free-standing 2D crystals are employed.

Dry-stamp transfer has emerged as a clean, reliable and efficient method for manipulation of mechanically exfoliated 2D flakes[33], but this approach is not possible using conventional TEM supports. Here we present a novel support grid that is compatible with direct dry-stamp transfer, facilitating preparation of ultra-clean large area TEM specimens for air-sensitive 2D materials and complex heterostructures (**Figure 1**). We begin with a standard $Si/SiN_x$ grid, which we chose to fabricate in-house to allow flexibility over the commercially available models. The $Si/SiN_x$ grid was prepared by patterning a $SiN_x/Si/SiN_x$ wafer using optical lithography and reactive ion etching (RIE) with a mixture of $CHF_3$ and $O_2$ to remove the exposed $SiN_x$. Following this, the sample was immersed in potassium hydroxide (KOH) to etch away the exposed bulk silicon leaving an isotropic etch profile as shown in **Figure 1a**. The sample was then flipped and the opposite $SiN_x$ membrane patterned with an array of holes using optical lithography and the same RIE process (**Figure 1b**). The size of these holes was typically 1-5 um diameter, providing a compromise between imaging area and specimen contact area (**Figure 1d**).

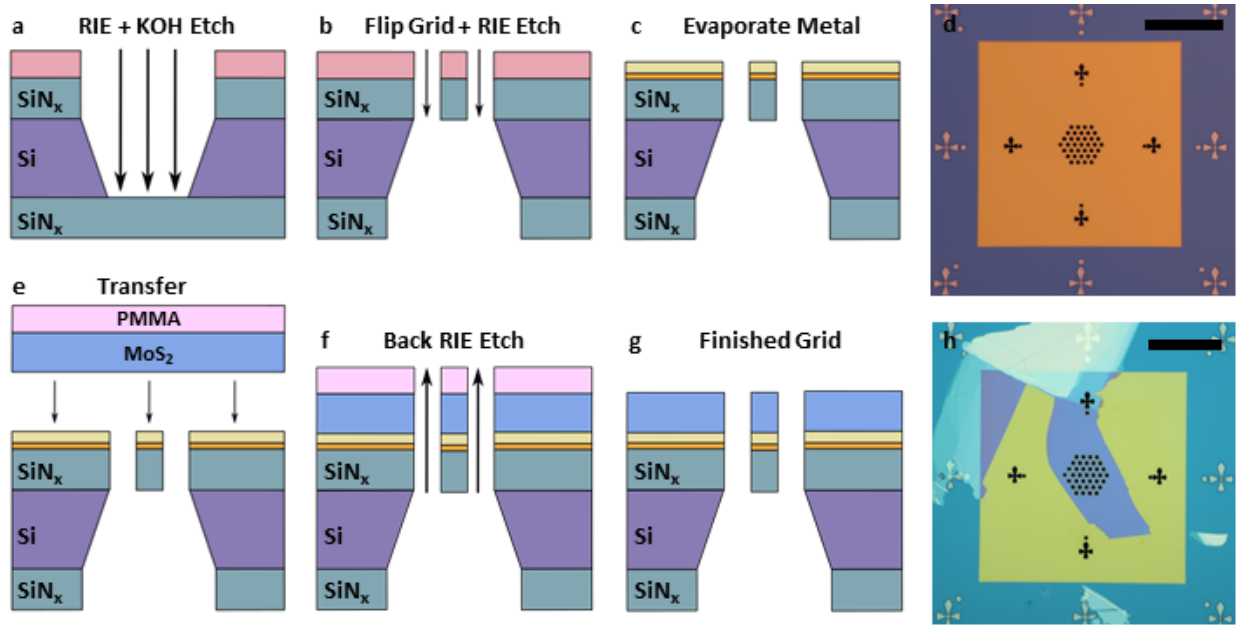

**Figure 1.** Fabrication procedure to create adhesion-enhanced $MoS_2$ grids. (a) First a $SiN_x$/Si/$SiN_x$ wafer is coated with photoresist (pink layer) then patterned using photolithography. Then the SiNx layer (teal) is removed by RIE and the exposed Si etched with KOH. (b) The grid is then flipped and the opposite side coated with photoresist. A grid pattern is written using optical lithography and RIE is then used to etch through the $SiN_x$. (c) Next Cr (1nm) and Au (5nm) adhesion layers are evaporated onto the grid with electron-beam evaporation. (d) Optical image of a patterned TEM grid. (e) In the next step a bulk crystal of mechanically exfoliated $MoS_2$ is transferred onto a grid using the PMMA dry-stamp transfer process. (f) The $MoS_2$ is back-etched through the grid. (g) The PMMA is dissolved in solvent and the grid is annealed at 450ºC in 10% $H_2$/Ar to remove surface residue. (h) Optical image of a finished adhesion-enhanced grid. Scale bar in (d) and (h) is 20μm.

To enhance this conventional $SiN_x$ grid design, a thin layer of chromium (1 nm) followed by gold (5 nm) was deposited onto the surface using electron-beam evaporation (**Figure 1c**). A large bulk $MoS_2$ crystal (>10 nm thickness, >50 μm x 50 μm in lateral area) was then transferred onto the grid using the poly(methyl-methacrylate) (PMMA) dry-stamp transfer technique[19] and a micromanipulation setup[34] (**Figure 1e**). The suspended regions of MoS$_2$ over the holes in the SiN$_x$ were then etched away using RIE from the back side of the grid with the SiN$_x$ serving as an etch mask (**Figure 1f**). Finally, the PMMA transfer layer was dissolved and the grid annealed at 450°C in 10% H$_2$/Ar for 6 hours (**Figure 1g**) to remove chemical residues and help bond the MoS$_2$ crystal to the SiN$_x$ by partial melting of the gold layer. We have found that annealing temperatures of at least 400°C are required to effectively bond the MoS$_2$, consistent with the expected melting point of thin Au films[35]. An optical image of the complete adhesion-enhanced grid (AEG) support is shown in **Figure 1h**. An extended description of the method is provided in **Supporting Information S1**.

A key advantage of our AEGs is their much lower surface roughness compared to SiN$_x$ grids or traditional holey carbon grids, achieved by the presence of the atomically smooth MoS$_2$. The roughness is greatly reduced at both the nanometer and micrometer length scales (**Figure 2**), which increases the van der Waals interaction between the grid and the 2D material or heterostructure specimen. Atomic force microscope (AFM) characterization of multiple areas on multiple grids gives an RMS roughness for our AEG grid of approximately 45 pm (the resolution limit of the AFM) compared to 320-450 pm for the SiN$_x$, and >800 pm for the carbon films tested. This improves the adhesion of atomically thin samples to the grids, reducing the probability of delamination or scrolling during removal of the polymer support film. Intrinsically better adhesion also removes the need for high temperature annealing as part of the transfer which, although it can improve specimen adhesion, is potentially damaging for heat sensitive samples. Importantly, the improved adhesion means our AEG supports are compatible with the dry-stamp mechanical transfer[33], which leaves significantly less polymer residue compared to conventional transfer[19]. This process relies on van der Waals attraction between the specimen and the substrate being stronger than between the specimen and the polymer support film, so that the specimen preferentially adheres to the substrate (see **Supporting Information S3**). We have found dry-transfer is not compatible with conventional TEM support grids: due to the small effective contact area the specimen remains attached to the polymer rather than the grid, or the polymer only partially delaminates causing damage to both the crystal and the grid.

We also tested the potential of other environmentally and thermally stable 2D materials for the 2D adhesion layer as alternatives to $MoS_2$. Hexagonal boron nitride (hBN) yielded similar positive transfer results, yet its insulating behavior made it unattractive for TEM studies where a conducting support helps to prevent specimen charging artefacts. Using graphite resulted in unwanted 'fencing' along the edges of the holes, providing additional roughness which interfered with subsequent transfers of the 2D crystal or heterostructure specimens (see **Supporting Information S4**). An additional benefit of using $MoS_2$ is its scalability, as few-layer uniform films produced by chemical vapor deposition (CVD) are becoming commercially available[36].

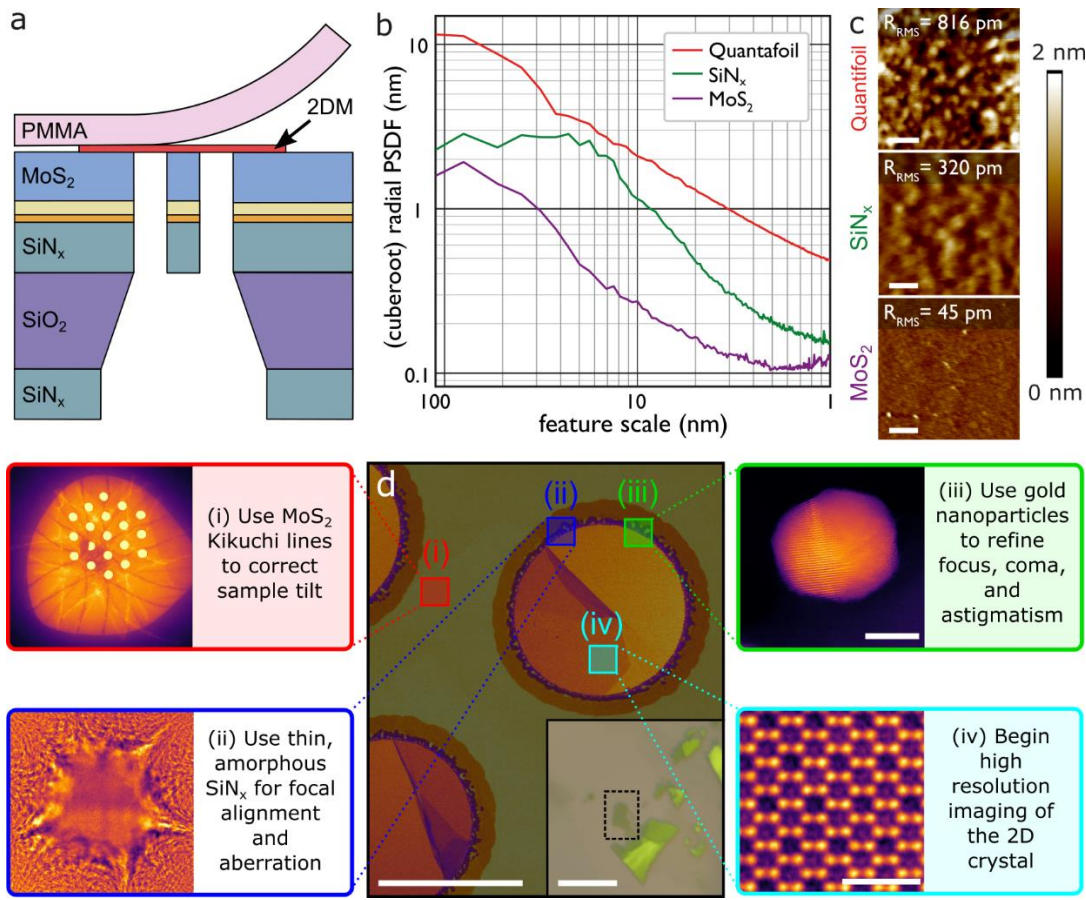

**Figure 2.** Sample transfer onto an AEG support. (a) Schematic of a 2D material dry-stamp transfer onto an AEG. The red layer is the transferred 2D specimen (2DM), the pink layer is the transfer polymer. (b) AFM roughness analysis comparing $MoS_2$, $SiN_x$ and Quantifoil TEM grids over different length scales and (c) surface roughness comparison taken over representative 500x500nm sampling areas. (d) False color STEM image of a 2D sample on an AEG, inset optical image of sample prior to AEG transfer (sample indicated by black dashed rectangle): (i-iv) a hierarchical approach to STEM imaging alignment using the structural features of an AEG. Scale bars (c) 100 nm, (d) 2 um, (d inset) 5 um, (iii) 5nm and (iv) 1nm.

In addition to the aforementioned benefits for sample fabrication, AEGs provide multiple practical advantages for effective high resolution TEM and scanning transmission electron microscopy

(STEM). Our developed hierarchical route to optimal STEM imaging is presented in **Figure 2d** where different components of the AEG are used to optimize the microscope set-up without compromising potential areas of interest. In brief the process is: (**i**) correct global sample tilt onto the desired [001] zone axis using the Kikuchi bands from the $MoS_2$ in the Ronchigram, (**ii**) set sample height and correct probe aberrations using thin, amorphous $SiN_x$ at the edge of the holes, (**iii**) fine-tune axial coma and astigmatism in the probe using Au nanoparticles in imaging mode, (**iv**) move to specimen area with optimized imaging conditions. Even in the latest generation of aberration corrected electron microscopes aberrations tend to drift from their optimal values, so optimal imaging conditions are only stable for minutes to hours[37]. The ability to correct for unwanted low order aberrations (e.g. defocus, astigmatism and axial coma) using the edge of the ultra-thin $SiN_x$ support and the Au nanoparticles, close to the region of interest is thus highly desirable, especially for the most electron beam sensitive 2D crystal samples. Furthermore, the very flat grid provides minimal variations in specimen tilt and height over distances of several microns, reducing the need to regularly reorient the sample and facilitating automated imaging procedures. The false-color STEM image is constructed using the method described in **Supporting Information S5**.

We have tested the efficacy of our AEGs on a wide range of 2D materials and heterostructure samples. We find that the overall success rate of specimen transfer is 97%, (2 failures from 67 sample transfers), compared to just 36% when transferring to uncoated $SiN_x$ TEM grids (39 failure from 58 samples, see **Table 1** for details). For both conventional grids and AEGs these results are based upon data collected from 10 individuals with experience of over 100 2D crystal transfers. Furthermore, the original uncoated $SiN_x$ grid transfer approach was already optimized based on many years' experience[38]. For example, we find that graphene encapsulation of air sensitive crystals increases the effective sample area, which increases the success rate of specimen transfer, in addition to the other previously reported benefits of charge dissipation[39] and use as a protective barrier[20]. However, the additional transfer steps associated with graphene encapsulation may also introduce additional contamination so is not always desirable. It is therefore most encouraging that the greatest improvement in transfer success rates is seen for laterally small 2D crystals or heterostructure stacks. Specimens with lateral sizes of ~ 5µm or below, such as twisted $MoS_2/WS_2$ bilayers[40], are indicated in **Table 1** and for these samples the success rate improves from 15% to 94% using the AEG approach.

**Table 1.** Transfer statistics comparing samples transferred onto SiNx grids and AEG supports. * indicates specimens with smaller areas due to the absence of graphene (Gr).

| Material | Attempts with $SiN_x$ Grid | Success rate with $SiN_x$ Grids (%) | Attempts with $MoS_2$ Grids | Success rate with $MoS_2$ Grids (%) |
|---|---|---|---|---|
| Gr | 10 | 60 | 10 | 100 |
| $Gr/CrBr_3/Gr$ | 4 | 50 | 5 | 100 |
| $Gr/CrGeTe_3/Gr$ | 4 | 50 | 5 | 100 |
| $Gr/InSe/Gr$[20] | 8 | 50 | 6 | 100 |
| $Gr/GaSe/Gr$[20] | 6 | 50 | 6 | 100 |
| ZIF-7* | 6 | 50 | 5 | 100 |
| Mica* | 10 | 10 | 10 | 90 |
| $MoS_2/WS_2$*[40] | 10 | 0 | 20 | 95 |
| All grids | 58 | 36 | 67 | 97 |

We illustrate the benefit of our AEGs by using them to enable STEM imaging of the air-sensitive magnetic 2D crystal $CrBr_3$ (see **Figure 3**). The few-layer specimen was prepared by micromechanical exfoliation in an inert argon glovebox[34], followed by graphene encapsulation, removal from the glovebox and dry-transfer on to a AEG. We have found that few-layer $CrBr_3$ crystals are sensitive to even moderately elevated temperatures thus annealing was avoided during the whole fabrication process. An extended description of the TEM/STEM methods is provided in **Supporting Information S6**.

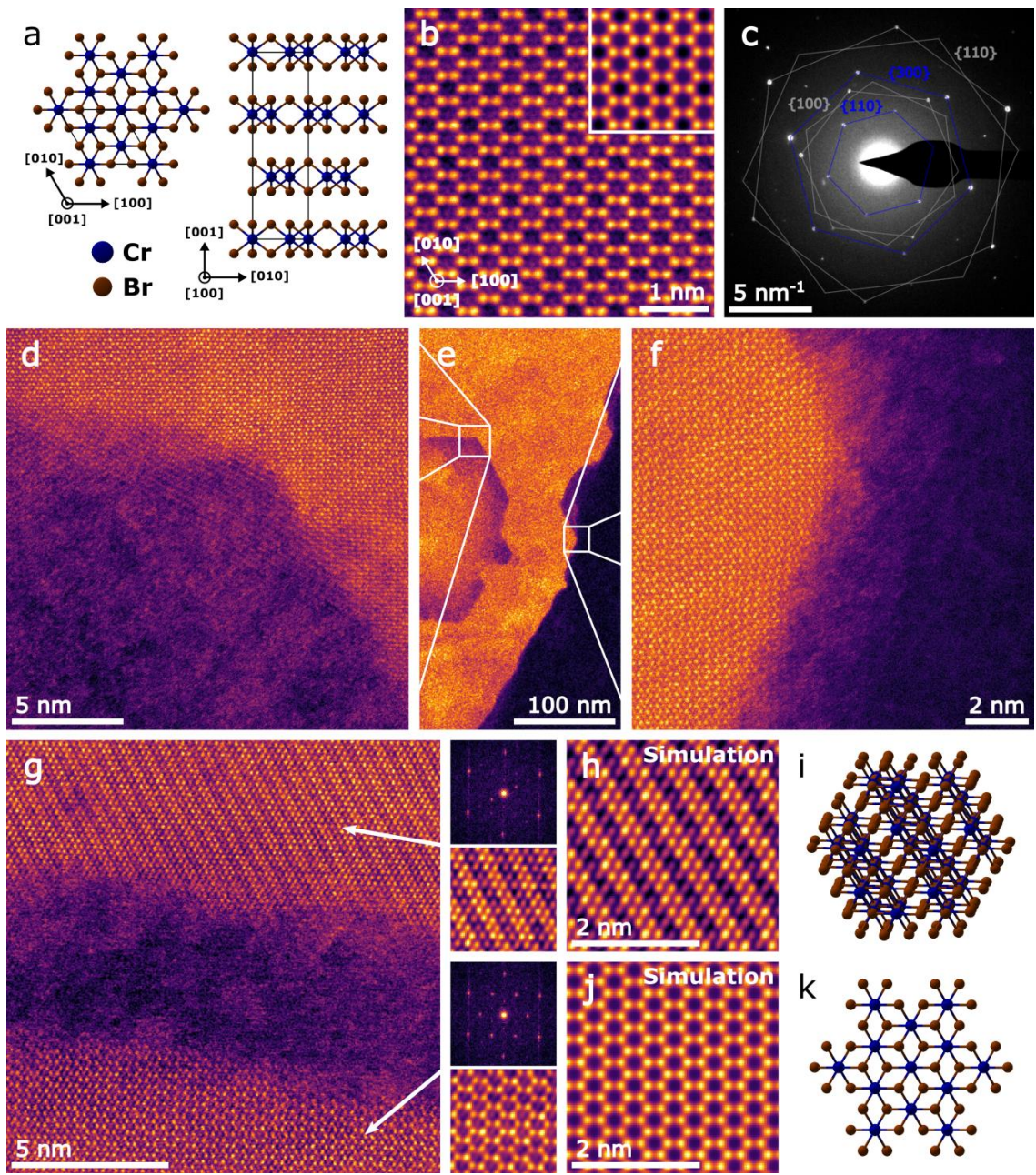

**Figure 3.** Atomic resolution ADF-STEM imaging of few-layer CrBr$_3$ crystals. (a) Model of trigonal CrBr$_3$, with unit cell dimensions indicated with black lines. (b) Drift-corrected, time-averaged, atomic resolution ADF-STEM image of 4-layer CrBr$_3$, showing strong agreement with image simulation (inset). (c) SAED pattern of a few-layer crystal with 1$^{st}$ and 2$^{nd}$ order spots for CrBr$_3$ (blue) and the graphene encapsulation (grey) indicated. (d – f) Faceted degradation observed in few-layer CrBr$_3$, with details (d, f) of the regions highlighted in (e). (g) 6-layer thick trigonal (lower) and monoclinic (upper) stacked regions of crystal separated by a crack, with Butterworth-filtered reduced FFTs (field-of-view: 14.5 nm$^{-1}$) and details (field-of-view: 2 nm) highlighting the difference in symmetry. (h-k) ADF-STEM image simulation and atomic model of 6-layer monoclinic (h,i) and trigonal (j,k) CrBr$_3$. The monoclinic stacking is observed at a small tilt (~ 2.5°) towards the [890] direction.

The specimens contained large, atomically clean areas suitable for high resolution (S)TEM imaging (**Figure 3**), which has not been reported previously for this class of material. Selected area electron diffraction (SAED) (**Figure 3c**) revealed successful preservation of the crystal structure at the micrometer scale *via* graphene encapsulation,[28,39]. The local thickness of the crystal was found to vary between 4 and 6 atomic layers, confirmed by comparison with atomic resolution image simulations as well as optical image contrast prior to encapsulation. The atomic number sensitivity of ADF-STEM imaging means that the two graphene encapsulation layers are virtually invisible and has negligible effect on image simulations where the elemental species of interest are significantly higher in atomic number than carbon[20]. Preventing structural degradation of CrBr$_3$ is a key concern for the application of this material so it is informative to consider what can be learned about this from STEM characterization of encapsulated samples. Even when graphene encapsulation is performed in an inert atmosphere, oxygen containing surface species are often trapped beneath the encapsulation layers and these tend to aggregate at edges of the specimen, in cracks or as pockets[41] on the surface. For air sensitive materials like $CrBr_3$ interaction with trapped oxidizing species causes local degradation as shown in **Figure 3e**. Interestingly, unlike in similar graphene encapsulated crystals of black phosphorus[42] and $TaS_2$[43], we observe that in $CrBr_3$, the boundaries of the degraded regions are highly faceted, preferring to terminate along the armchair directions (the $\{100\}$ and $\{010\}$ planes of the crystal), seen both at holes formed in the center of the crystal (**Figure 3d**) and at edges (**Figure 3f**). As holes in the center of a crystal are attributed to electron beam induced reaction of $CrBr_3$ with oxidizing species in trapped contamination pockets, their faceting confirms the phenomena is degradation induced, rather than a property of the crystal growth or of the mechanical exfoliation process. Although, the edge terminations are clearly crystallographically defined at the nanoscale, atomic resolution imaging suggests that the faceting is not atomically sharp. To further investigate the edge structure, we have performed STEM elemental mapping using electron energy loss and energy dispersive x-ray spectroscopy (EELS/EDXS) (see **Supporting Information S7**). This revealed a decrease in Br concentration relative to pristine $CrBr_3$, with a Cr:Br ratio of 1:2.5 at the edge versus 1:3 in the center, suggesting that Br is preferentially lost from the crystal and leaving Br vacancies in the edge region. The loss of Br is not unexpected as this element is known to readily intercalate into graphite[44] and therefore can easily diffuse away between the encapsulating graphene sheets. Unexpectedly we detect no significant oxygen layer associated with the crystal edges, which is in strong contrast with what is observed for other graphene encapsulated air sensitive crystals[42,43]. Instead we observe only a carbon rich layer at the edge of the crystal, which is likely to be associated with a small amount of residual polymer contamination. We speculate that this could play a role in stabilizing the Br vacancies and facetted edges, although first principles calculations would be required to better understand this behavior.

Our atomic resolution ADF-STEM imaging reveals the expected trigonal crystal stacking in the majority of the sample (**Figure 3b**)[45-47]. However, in **Figure 3g**, the stacking is seen to transition from the trigonal sequence (lower region) to a monoclinic sequence (upper region), with the two separated by a crack in the crystal. The differences in crystal symmetry are highlighted in the inset reduced Fast Fourier Transforms (rFFTs), where the change in stacking affects the intensity of the inner spots. This change in stacking is confirmed by comparison with image simulations (**Figure 3h**, **j**), which reveals that the stripe contrast is caused by the alignment of mixed Cr – Br (higher total Z, bright stripes) and Br – hollow site (lower total Z, dark stripes) atomic columns in this stacking. This sequence is similar to previously reported monoclinic $CrX_3$ (X = Cl, Br, or I)[48,49] (an atomic model is presented in **Supporting Information S8/S9,** where we also compare the change in stacking to the strip contrast that can be produced by crystal tilt alone). Although local stacking changes can result during synthesis of 2D materials, the transition from trigonal to monoclinic can be achieved solely through shear forces, and the proximity of this structure to the crack suggests that here the transition is likely to be due to mechanical deformation resulting in the local slippage of atomic layers on one side of the crack.

**Conclusion**

In summary, we present a new design of adhesion-enhanced support grids that facilitate the clean and reliable preparation of suspended samples for 2D crystals and their complex heterostructures by the dry-stamp transfer method. These AEGs supports can be produced in bulk using traditional clean room fabrication processes and production could be scaled up to wafer scale through the use of CVD MoS$_2$. We find the specimen preparation approach facilitated by the use of such grids has a 100% success rate for graphene encapsulated samples and 94% success rate for 2D crystals and heterostructures with small lateral dimensions. This much improved reliability compared to conventional support grids is invaluable when preparing free-standing specimen from the most challenging air-sensitive 2D crystals and sophisticated heterostructures. We demonstrate the successful application of our technique by atomic resolution STEM imaging of few-layer CrBr$_3$ films, revealing a range of new local structural detail including facetted crystal degradation and mechanically induced stacking faults.

**Supporting Information**. **S1** describes the adhesion-enhanced grid fabrication methodology. **S2** discusses scrolling effects on low adhesion TEM grids. **S3** demonstrates stamp transfer onto adhesion-enhanced grids. **S4** compares alternative materials for TEM grids. **S5** provides low magnification STEM data. **S6** displays CrBr$_3$ EELS/EDX data. **S7** describes the TEM/STEM methodology. **S8** and **S9** provide additional details about the TEM investigations.

## AUTHOR INFORMATION

**Corresponding Author**

Roman Gorbachev

Roman@Manchester.ac.uk

Sarah Haigh

Sarah.Haigh@Manchester.ac.uk

**Author Contributions**

R.G. and S.J.H conceived the study; M.Z fabricated devices with help from M.H, N.C, T.B and W.W. TEM investigations of the samples were performed by D.G.H, D.K and Y.Z. The manuscript was written by M.H, N.C, D.G.H, S.J.H and R.G with contributions from all authors.

**Funding Sources**

We acknowledge support from EPSRC grants EP/N509565/1, EP/P01139X/1, EP/N010345/1 and EP/L01548X/1 along with the CDT Graphene-NOWNANO, and the EPSRC Doctoral Prize Fellowship. We thank Diamond Light Source for access and support in use of the electron Physical Science Imaging Centre (Instrument E02 and proposal numbers EM19315 and MG21597) that contributed to the results presented here. In addition, we acknowledge support from the European Commission including H2020 program grants: European Graphene Flagship Project (696656), European Quantum Technology Flagship Project 2DSIPC (820378), ERC Synergy Grant Hetero2D and ERC Starter grant EvoluTEM (715502).

**Notes**

Additional data related to the paper may be requested from the authors. The authors declare no competing financial interests.